\documentclass[aps,prd,twocolumn,showpacs,groupedaddress]{revtex4} 
\usepackage{graphicx}  
\usepackage{dcolumn}   
\usepackage{bm}        
\usepackage{amssymb}   
\usepackage{amsmath}

\begin{document}

\title{Gravitational Wave Burst Source Direction Estimation using Time and Amplitude Information}
\author{
J. Markowitz,
M. Zanolin, 
L.Cadonati, 
and E. Katsavounidis
\\
}
\affiliation{
\centerline{LIGO - Massachusetts Institute of Technology, Cambridge, MA 02139 USA}
}
\date{\today}

\begin{abstract}
In this article we study two problems that arise when using timing and amplitude estimates from a network of interferometers (IFOs) to evaluate the direction of an incident gravitational wave burst (GWB).  First, we discuss an angular bias in the least squares timing-based approach that becomes increasingly relevant for moderate to low signal-to-noise ratios.  We show how estimates of the arrival time uncertainties in each detector can be used to correct this bias. We also introduce a stand alone parameter estimation algorithm that can improve the arrival time estimation and provide root-sum-squared strain amplitude ($h_{rss}$) values for each site.  In the second part of the paper we discuss how to resolve the directional ambiguity that arises from observations in three non co-located interferometers between the true source location and its mirror image across the plane containing the detectors.  We introduce a new, exact relationship among the $h_{rss}$ values at the three sites that, for sufficiently large signal amplitudes, determines the true source direction regardless of whether or not the signal is linearly polarized.  Both the algorithm estimating arrival times, arrival time uncertainties, and $h_{rss}$ values and the directional follow-up can be applied to any set of gravitational wave candidates observed in a network of three non co-located interferometers.  As a case study we test the methods on simulated waveforms embedded in simulations of the noise of the LIGO and Virgo detectors at design sensitivity.
\end{abstract}

\maketitle

\section*{I.  Introduction}

  As the first generation of gravitational wave interferometers perform observations at or near their design sensitivities, new methods are being developed to detect and characterize gravitational wave burst signals.  Accurate determination of the source direction is fundamental for all analyses of a candidate signal.
  
  There are two approaches to localizing a source on the celestial sphere: coincident and coherent.  Coincident methods\cite{fabien} analyze the data from each detector separately and then identify events that occur simultaneously in multiple interferometers.  The arrival times and amplitudes in each detector can then be used to determine the source direction if the signal is linearly polarized or has polarization peaks separated by less than the timing uncertainty.  Coherent methods \cite{gt,scott,sergey,patrick} combine the data streams from multiple detectors into a single statistic before reconstructing gravitational wave candidates.

  One previous coincident study on source location estimation was completed by Cavalier et al\cite{fabien}.  In that work, arrival times (assumed known) and their uncertainties (assumed gaussian) were input to a $\chi^2$ minimimization routine to determine the source location.  The approach is applicable to an arbitrary number of interferometers and shows excellent resolution in Monte Carlo simulations with arrival time uncertainties on the order of 0.1 ms.  In this paper we implement an equivalent least squares approach in the two and three interferometer cases, but also consider arrival time uncertainites up to 3 ms.  These larger uncertainties have been observed in studies with real noise\cite{aps,lvburst} and reveal a systematic bias in the source directions obtained from timing reconstruction methods for moderate to low signal-to-noise ratios (SNRs).  We define the SNR $\rho$ as 

\begin{equation}
\rho \equiv \sqrt{4\int_0^\infty \frac{|\tilde{h}(f)|^{2}}{S(f)}\,df},
\end{equation}

\noindent where 

\begin{equation}
\tilde{h}(f) = \int_{-\infty}^\infty h(t)e^{-i2\pi ft} dt
\end{equation}
 
is the Fourier Transform of the gravitational wave signal and 

\begin{equation}
S(f) = |\tilde{n}(f)|^2 + |\tilde{n}(-f)|^2 \;\;\;\;\;\;\; (0\le f \le \infty) 
\end{equation}

 is the one-sided power spectrum of the detector noise.  We will show that the bias is most significant near the plane of interferometers in the three detector case, agreeing with the observation of reduced resolution in this region by \cite{fabien}.  We compute a numerical table of corrections for the three-interferometer bias and note that a similar procedure may be followed for an arbitrary network of detectors.  We also find that the bias may be corrected through the application of a more detailed parameter estimation \cite{michele}, which has the effect of reducing arrival time errors.  Our application of the least squares estimator with bias correction to simulated data is the first test of coincident source localization with simulated detector noise.

  Coherent methods also offer promise for source direction estimation.  The first coherent analysis was described by G\"{u}rsel and Tinto\cite{gt}.  Their approach combines the data from a three interferometer network to form a ``null stream'' where the gravitational wave signal should be cancelled completely if the assumed source direction is correct.  The true source direction is estimated by looping over a grid of angular locations and minimizing the result.  Other grid-based approaches to source localization include the maximum likelihood approach of Flanagan and Hughes\cite{scott} and the constraint likelihood method of Klimenko et al\cite{sergey}.  The required resolution of the grids used in these analyses increases with the maximum frequency of the search; for LIGO burst searches that extend past $1000$ Hz\cite{ligo3} approximately $10^6$ grid points are necessary.  In practice much coarser grids must be used due to computational limitations, potentially leading to missed minima and incorrect directional estimates.  It should also be noted that coherent methods implemented with constraints are insensitive to some signal morphologies and (small) sky regions.
   
  The amplitude test described here is similar to the null stream approach, but does not require minimization over a grid because it works in terms of the root-sum-squared strain amplitude 

\begin{equation}
h_{rss} \equiv \sqrt{\int h(t)^2\,dt}.
\end{equation}

\noindent Throughout this paper $h_{rss}$ will refer to the ``intrinsic'' gravitational wave signal at the Earth, i.e. prior to reduction by the detector antenna pattern.  The test uses the $h_{rss}$ values from three non co-located interferometers to choose between the two possible source directions given by timing considerations and is effective for moderate to high SNR. 

  The organization of this paper is as follows.  Section II describes the angular bias in the least squares approach to source localization using arrival times from either two or three interferometers.  In each case, a Monte Carlo simulation is pursued and used to characterize and numerically correct the observed bias.  Section III introduces the aforementioned amplitude check that can be used to differentiate between the two possible source locations given by arrival time considerations in the three interferometer geometry.  Both the time-based source direction estimator in (II) and the amplitude check in (III) were tested on LIGO-Virgo simulated data and the results are described in Section IV.

\section*{II. Least Squares Source Localization with Bias Correction}
In the following we examine the standard least squares approach to source localization for both the two and three interferometer geometries.  We restrict ourselves to studying short duration ($<$ 1 s) ``burst'' signals so that the motion of the Earth may be neglected.  We also assume that the difference in travel time between sites is due only to the direction of the source, as explained in the introduction. 

\begin{figure}
\includegraphics[width=0.45\textwidth]{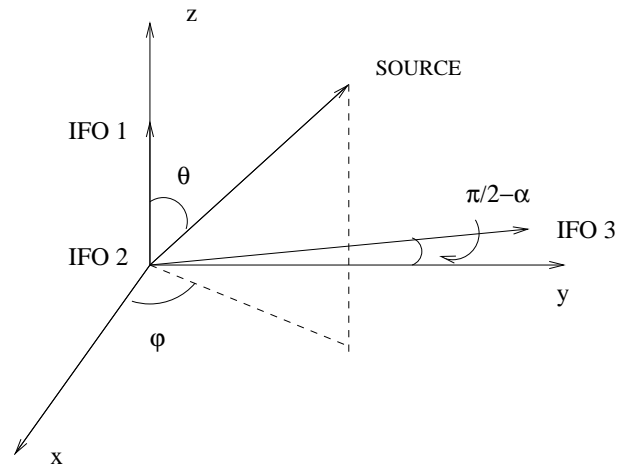}
\caption{Coordinate system used for source localization.  Only the polar angle $\theta$ is constrained in the two interferometer (IFO) case, while both angles $\theta$ and $\phi$ are constrained in the three IFO case.}
\end{figure}

\subsection*{A.  Two Interferometer Case}
     When the network consists of two detectors at different locations, the angle $\theta$ between the source unit vector and the vector connecting the sites (baseline) is given by 
\begin{equation}
\theta = \cos^{-1}\left( \frac{c\tau}{d} \right),
\end{equation}
where $\tau$ is the time delay, $d$ is the distance between detectors, and $c$ is the speed of light. This relationship is only exact in the absence of noise and constrains the source direction to a ring on the sky.  When the arrival time estimates are affected by noise, the following least squares estimator can be used:

\begin{equation}
\hat{\theta}= 
\begin{cases} \pi &\text{for $\tau < -d/c$,}\\
\cos^{-1}(\frac{c\tau}{d}) &\text{for $-d/c \le \tau \le d/c$,}\\
0 &\text{for $\tau > d/c$.}
\end{cases}
\end{equation}

\noindent This is the Maximum Likelihood Estimator and is the optimal choice for small arrival time errors.  Note that the addition of noise introduces timing errors that may lead to estimated time delays greater than the light travel time between detectors.  These unphysical delays are mapped to the polar angle that produces the time delay closest to what is observed.

\begin{figure}
\includegraphics[width=0.45\textwidth]{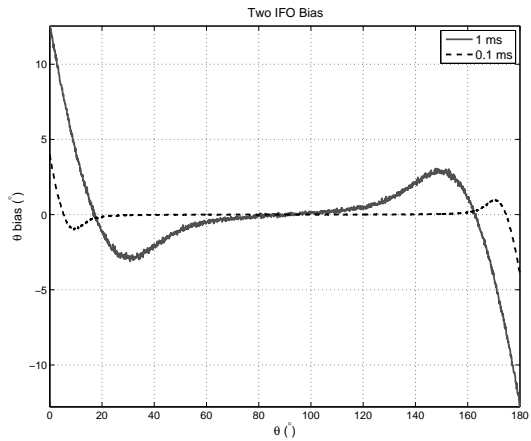}
\caption{Least squares angular bias for the Livingston-Hanford (L1-H1) baseline with a gaussian arrival time error distribution and $\sigma_{t_H}$= $\sigma_{t_L}$ = 0.1 ms, 1 ms.}
\end{figure}

\begin{figure}
\includegraphics[width=0.45\textwidth]{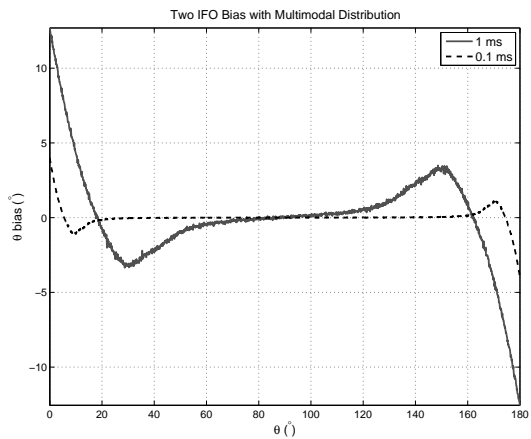}
\caption{Least squares angular bias for the Livingston-Hanford (L1-H1) baseline with a multimodal arrival time error distribution and $\sigma_{t_H}$= $\sigma_{t_L}$ = 0.1 ms, 1 ms.}
\end{figure}

   The characteristics of the estimator (6) were studied through multiple Monte Carlo simulations.  In the first simulation we assumed that the distribution of measured arrival times for a given source location and detector follows a gaussian distribution with mean equal to the true arrival time and variance equal to the variance of the arrival time estimate.  In the second simulation we assumed that the distribution of measured arrival times for a given source location and detector follows a multimodal distribution consisting of a gaussian main lobe and two side lobes each containing half as many points as the main lobe.  This second scenario is often the case for signals with multiple peaks. In both cases we chose coordinates as indicated in Figure 1 and used the time delay provided by the LIGO Livingston (L1) - LIGO Hanford (H1) baseline.  Twenty thousand sets of arrival times (with simulated error) from 1800 different sky positions were produced via a gaussian random number generator.  The sky points were spaced by $0.1^\circ$ on the $\phi$ = $0^\circ$ meridian of our symmetric coordinate system.  For each sky location, the estimator expectation value $<\hat{\theta}>$ and its variance $\sigma_{\hat{\theta}}^2$ were computed.  A nonzero systematic bias 
\begin{equation}
B_{\hat{\theta}} = <\hat{\theta}> - \theta
\end{equation}
was observed and is plotted against true polar angle $\theta$ in Figures 2 and 3.  The bias in each trial is similar and in each case increases with the variance of the arrival time distribution.  We observed the bias to grow with the arrival time uncertainty, and expect it to maintain a similar shape for symmetric distributions.  The bias is a result of two features of the estimator:

1.)  Mapping of unphysical time delays ($|\tau| > d/c$) to the two interferometer baseline has the effect of pushing the expectation value of the estimated angle towards the equator.

2.)  The nonlinear mapping between the time delay $\tau$ and the angle $\theta$ contributes to the bias through the following relation:

\begin{equation}
|p(\hat{\theta})d\hat{\theta}| = |p(\tau)d\tau|.
\end{equation}

This implies that the distribution $p(\hat{\theta})$ will reflect the distribution of time delays $p(\tau)$ modulated by a sinusoidal Jacobian.  Assuming a normal distribution for $p(\tau)$ results in the bias shape seen in Figure 2, while the multimodal distribution discussed above yields the shape seen in Figure 3.  Note that the bias is a result of the network geometry and is therefore independent of coordinate choice.  We verified this statement numerically by observing the same bias for every sky position when $\theta$ was chosen along an axis different from the interferometer baseline.  We also verified this property in the three interferometer case, as discussed below.

\subsection*{B. Three Interferometer Case}

  In the three interferometer case there are two independent baselines, so a least squares estimator will constrain the source direction to two patches on the sky.  These patches will be mirror images around the plane formed by the three detectors and are indistinguishable because they yield identical time delays.  The plane of interferometers is the y-z plane in our chosen coordinate system as shown in Figure 1.  With these coordinates, we can easily determine the output of a least squares estimator given three arrival times.  The delay times $\tau_1$, $\tau_2$ and their uncertainties are defined in terms of the three arrival times $t_1$, $t_2$, $t_3$:

\begin{equation}
\tau_1 \equiv t_2-t_1,
\end{equation}
\begin{equation}
 \tau_2 \equiv t_2-t_3, 
\end{equation}
\begin{equation}
\sigma_{\tau_1} = \sqrt{\sigma_{t_1}^2 + \sigma_{t_2}^2},
\end{equation}
\begin{equation}
\sigma_{\tau_2} = \sqrt{\sigma_{t_2}^2 + \sigma_{t_3}^2}.
\end{equation}

The uncertainties were defined assuming a gaussian spread around each true arrival time.  Using these definitions the least squares location estimator (also the maximum likelihood estimator if the arrival times follow a gaussian distribution) is equivalent to the following:

1.)  If $|\tau_1| > d_1/c$ \& $|\tau_2| > d_2/c$:

\begin{equation}
(\hat{\theta}, \hat{\phi})=
\begin{cases} (0, \pi) &\text{if  $\frac{\tau_1-d_1/c}{\sigma_{\tau_1}} > \frac{|\tau_2|-d_2/c}{\sigma_{\tau_2}}$,}\\
(\pi, \pi) &\text{if  $\frac{-\tau_1-d_1/c}{\sigma_{\tau_1}} > \frac{|\tau_2|-d_2/c}{\sigma_{\tau_2}}$,}\\
(\alpha, \pi/2) &\text{if  $\frac{\tau_2-d_2/c}{\sigma_{\tau_2}} > \frac{|\tau_1|-d_1/c}{\sigma_{\tau_1}}$,}\\
(\pi-\alpha, 3\pi/2) &\text{if  $\frac{-\tau_2-d_2/c}{\sigma_{\tau_2}} > \frac{|\tau_1|-d_1/c}{\sigma_{\tau_1}}$,}\\
\end{cases}
\end{equation}

2.)  If $|\tau_1| > d_1/c$ \& $|\tau_2| \le d_2/c$:

\begin{equation}
(\hat{\theta}, \hat{\phi})=
\begin{cases} (0, \pi) &\text{if $\tau_1 \ge d_1/c$,}\\
(\pi, \pi) &\text{if $\tau_1 \le -d_1/c$}\\
\end{cases}
\end{equation}

3.)  If $|\tau_1| \le d_1/c$ \& $|\tau_2| > d_2/c$:

\begin{equation}
(\hat{\theta}, \hat{\phi})=
\begin{cases} (\alpha, \pi/2) &\text{if $\tau_2 > d_2/c$,}\\
(\pi-\alpha, 3\pi/2) &\text{if $\tau_2 < -d_2/c$}\\
\end{cases}
\end{equation}

4.)  If $|\tau_1| \le d_1/c$ \& $(d_2/c)(\sin(\hat{\theta})\sin(\alpha)+\cos(\hat{\theta})\cos(\alpha)) < \tau_2 \le d_2/c$:

\begin{equation}
(\hat{\theta}, \hat{\phi})=\left(\cos^{-1}\left(\frac{c\tau_1}{d_1}\right), \pi/2\right)
\end{equation}

5.)  If $|\tau_1| \le d_1/c$ \&  $-d_2/c \le \tau_2 < (d_2/c)(-\sin(\hat{\theta})\sin(\alpha)+\cos(\hat{\theta})\cos(\alpha))$:

\begin{equation}
(\hat{\theta}, \hat{\phi})=\left(\cos^{-1}\left(\frac{c\tau_1}{d_1}\right), 3\pi/2\right)
\end{equation}

6.)  If $|\tau_1| \le d_1/c$ \& $(d_2/c)(\sin(\hat{\theta})\sin(\alpha)+\cos(\hat{\theta})\cos(\alpha)) \le \tau_2 \le (d_2/c)(-\sin(\hat{\theta})\sin(\alpha)+\cos(\hat{\theta})\cos(\alpha))$:

\begin{equation}
\hat{\theta} = \cos^{-1}\left(\frac{c\tau_1}{d_1}\right),
\end{equation}
\begin{equation}
\hat{\phi}=\sin^{-1}\left( \frac{c\tau_2/d_2 - \cos(\hat{\theta})\cos(\alpha)}{\sin(\hat{\theta})\sin(\alpha)} \right).
\end{equation}

The case (1) occurs when both time delays are unphysical and pins the source location on the baseline that is the most standard deviations from physical.  The cases (2) and (3) occur when one time delay is unphysical and places the source location along the appropriate baseline.  The cases (4) and (5) are a result of analyzing one angle first, and place the source at the correct azimuthal angle if the second time delay is unphysical given the first.  The last case is the only one where both time delays together yield a physical result.  Note that the choice of $\hat{\phi}$ is arbitrary when the source is located on a pole.

\begin{figure}
\includegraphics[width=0.45\textwidth]{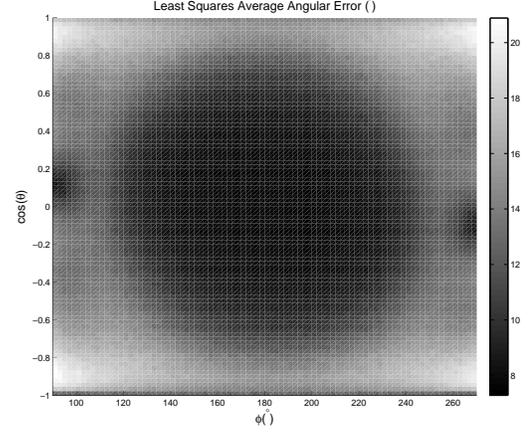}
\caption{Average angular error in location reconstruction from the least squares method applied to the H1-L1-V1 network when $\sigma_{t1}$ = $\sigma_{t2}$ = $\sigma_{t3}$ = 1 ms.  Note that only one hemisphere is shown here as we assume the correct sky patch is chosen and the result in the other atmosphere will be the mirror image of this plot across the y-z plane.}
\end{figure}

\begin{figure}
\includegraphics[width=0.45\textwidth]{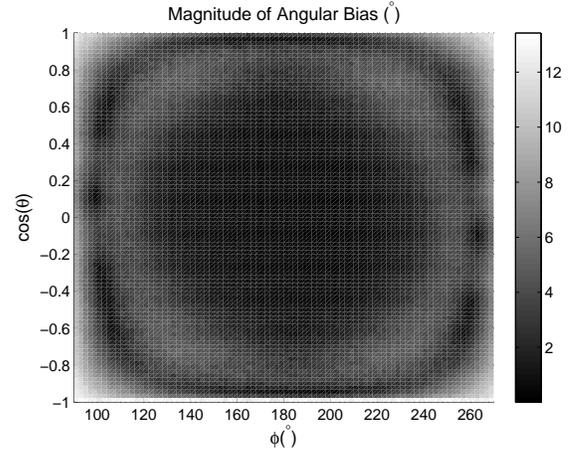}
\caption{Magnitude of 3 IFO (Hanford (H1), Livingston (L1), Virgo (V1)) angular bias when $\sigma_{t1}$ = $\sigma_{t2}$ = $\sigma_{t3}$ = 1 ms.  This magnitude is defined by (22), and the color scale is in degrees. Note that only one hemisphere of the sky is shown here, as the bias of the other hemisphere is its mirror image across the y-z plane.}
\end{figure}	

\begin{figure}
\includegraphics[width=0.45\textwidth]{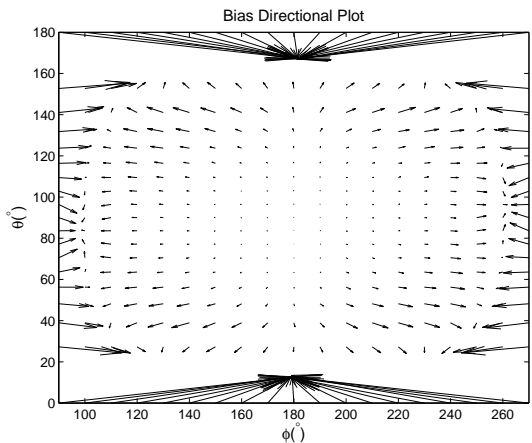}
\caption{Direction of 3 IFO (H1, L1, V1) angular bias when $\sigma_{t1}$ = $\sigma_{t2}$ = $\sigma_{t3}$ = 1 ms.}
\end{figure}	



  The performance of this least squares analysis was evaluated through a Monte Carlo simulation similar to that used for the two interferometer case and similar to those conducted in \cite{fabien}.  The simulation consisted of 8281 grid points spread isotropically across one hemisphere.  The hemisphere was chosen to be $(\pi/2 \le \phi \le 3\pi/2,0 \le \theta \le \pi)$ in our symmetric coordinates (Figure 1); the results for the other hemisphere would be a mirror image of these.  For 3000 iterations at each grid point, the true arrival time in each detector was added to normally distributed random noise.  This gaussian noise was taken to have zero mean and variance equal to that of the arrival time.

  The average angular error in the location reconstruction of this simulation is shown in Figure 4.  This error was calculated by first computing the angle between the true source position (grid point) and estimated location in each trial.  These angular errors were then averaged for each true position (grid point) and turned into the map shown in Figure 4.  As in the two interferometer case, a systematic bias made a significant contribution to the overall error.  We define the bias in the polar and azimuthal angles as
\begin{equation}
B_{\hat{\theta}} = <\hat{\theta}> - \theta
\end{equation}
and
\begin{equation}
B_{\hat{\phi}} = <\hat{\phi}> - \phi,
\end{equation}
respectively.  Note that trials where the source was placed on a pole were only counted toward the expectation value $<\hat{\theta}>$, as $\theta$ is well-defined at these points but $\phi$ is not.  The total magnitude of the angular bias was calculated using all trials and is given by

\begin{equation}
M = \cos^{-1}(\vec{P} \cdot \vec{Q}), 
\end{equation}

\noindent where $\vec{P}$ is the unit vector in the direction $(<\hat{\theta}>, <\hat{\phi}>)$ and $\vec{Q}$ is the unit vector in the direction of the source. A skymap of the value of $M$ for the Hanford-Livingston-Virgo (H1-L1-V1) network is plotted in Figure 5, while Figure 6 shows the direction of the bias.  An arrival time uncertainty of 1 ms was assumed for all detectors in both plots; in general the bias increases as arrival time uncertainty increases. 

  Note that the location estimation performed by this algorithm and the bias correction are independent of coordinate choice.  This is because the estimator places the source in the bin that yields time delays closest to what is observed regardless of whether the input time delays are physical.  We verified this property by recomputing the bias in an arbitrary coordinate system and observing the same results upon rotation back to our original coordinates. 

  The bias observed in the three interferometer geometry can be easily corrected numerically.  For a given network configuration and set of arrival time uncertainties, one can construct a sky map of the effect of the bias.  If the output of the least squares estimator is then assumed to be equal to the expectation value of the estimator, the output coordinates can be matched to a term in the bias array and the effect of the bias subtracted out.  As evidenced by Figures 4 and 5, this correction can reduce the mean square errors on the estimated angles significantly.  The numerical bias correction was applied to LIGO-Virgo simulated data and its performance is described in Section IV. 





\section*{III.  Amplitude Constraints}
 Measurements of the gravitational wave signal amplitude at different sites in a network provide valuable information for source localization.  In the following we will describe an amplitude check that resolves the source direction degeneracy inherent in the three interferometer timing analysis, provided the signal-to-noise ratio is sufficiently high in each detector.  The antenna patterns will be calculated using the method of Thorne \cite{300} as presented by Yakushin, Klimenko, and Rakhmanov\cite{ykr}.  \\

\subsection*{A. Detector Response}
     In the source frame, the gravitational wave metric $\mathbf{h_{\mu\nu}}$ can be written in the transverse traceless gauge as
\begin{equation}
\mathbf{h_{\mu\nu}}(t) =  \left( \begin{array}{cccc}
0 & 0 & 0 & 0\\
0 & h_{+}(t) & h_{\times}(t) & 0\\
0 & h_{\times}(t) & -h_{+}(t) & 0\\
0 & 0 & 0 & 0 \end{array} \right)
\end{equation} 
where $h_{+}(t)$ and $h_{\times}(t)$ are the two independent polarizations of the source.  For gravitational wave bursts of short duration, the time dimension can be taken to be fixed and we can define the traceless and transverse vectors
\begin{equation}
\textbf{m} = \left( \begin{array}{ccc}
1 & 0  & 0 \\
0 & -1 & 0 \\
0 & 0  & 0 \end{array} \right)
\end{equation}
and
\begin{equation}
\textbf{n} = \left( \begin{array}{ccc}
0 & 1 & 0 \\
1 & 0 & 0 \\
0 & 0 & 0 \end{array} \right).
\end{equation}

\begin{figure}
\includegraphics[width=0.45\textwidth]{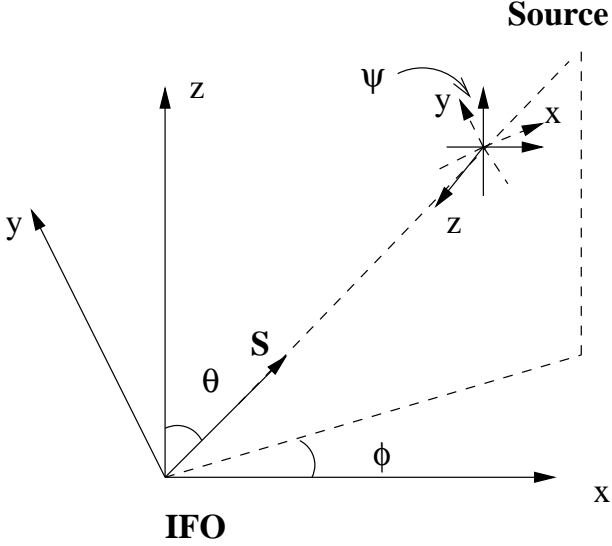}
\caption{Coordinate Transformation from IFO to GW frame.  The IFO frame is defined naturally by the x and y arms of the detector.  Figure is adopted from \cite{ykr}.}
\end{figure}

\noindent The strain tensor is then
\begin{equation}
\textbf{h}(t) = h_{+}(t)\textbf{m} + h_{\times}(t)\textbf{n}.
\end{equation}
The signal in an interferometer due to an incident gravitational wave is given by
\begin{equation}
S(t) = F_{+}h_{+}(t) + F_{\times}h_{\times}(t),
\end{equation}
where $F_{+}$ and $F_{\times}$ are the antenna patterns of the detector with respect to the two polarizations.  As shown in \cite{ykr}, these coefficients are equal to
\begin{equation}
F_{+} = \frac{1}{2}\text{Tr}(\mathbf{mR^{T}mR})
\end{equation} 
and
\begin{equation}
F_{\times} = \frac{1}{2}\text{Tr}(\mathbf{mR^{T}nR}),
\end{equation} 
where $\textbf{R}$ is the rotation from the interferometer frame to the source frame.  This rotation matrix may be decomposed into three Euler rotations, as shown in Figure 7.  The total rotation is thus
\begin{equation}
\textbf{R} = \mathbf{R_z}(\psi - \pi/2)\mathbf{R_y}(\theta - \pi)\mathbf{R_z}(\phi),
\end{equation}
where $\psi$ is the polarization angle. Now note that
\begin{equation}
\mathbf{R_{z}^{T}}(\psi)\mathbf{hR_z}(\psi) = \mathbf{h'},
\end{equation} 
where $\mathbf{h'}$ has the same form as $\mathbf{h}$ with
\begin{equation}
h'_{+}(t) = \cos(2\psi)h_{+}(t) + \sin(2\psi)h_{\times}(t)
\end{equation}
and
\begin{equation}
h'_{\times}(t) = -\sin(2\psi)h_{+}(t) - \cos(2\psi)h_{\times}(t).
\end{equation}
Since the transverse traceless form of $\mathbf{h}$ is preserved by the final Euler rotation, we may absorb the polarization angle into the choice of ($h_{+}(t),h_{\times}(t)$).  This choice will not affect the response of any interferometer to the gravitational wave source, but does allow us to write new antenna patterns as a function of only the source direction. 
\subsection*{B. Amplitude Check}

\begin{figure}
\includegraphics[width=0.45\textwidth]{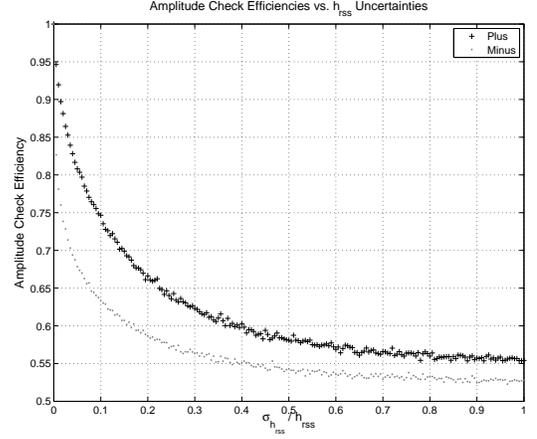}
\caption{Efficiency of the two amplitude checks plotted as a function of the uncertainty in recovered $h_{rss}$ values.  The plus check is given by (40) while the minus check is given by (41).}
\end{figure}

The response of a three interferometer network to gravitational radiation from direction ($\theta,\phi$) will have the following form:
\begin{equation}
h_{1}(t) = F_{+,1}h_{+}(t) + F_{\times,1}h_{\times}(t)
\end{equation}
\begin{equation}
h_{2}(t + \Delta t_{12}) = F_{+,2}h_{+}(t) + F_{\times,2}h_{\times}(t)
\end{equation}
\begin{equation}
h_{3}(t + \Delta t_{13}) = F_{+,3}h_{+}(t) + F_{\times,3}h_{\times}(t),
\end{equation}
Here the $h_{i}$ terms are the gravitational wave signals, $\Delta t_{12}$ and $\Delta t_{13}$ are the time delays from IFO 1 to IFO 2 and IFO 3, and $F_{+,i}$ and $F_{\times,i}$ are the plus and cross antenna patterns in each detector.  Note that the time delays are a function of the source direction and we have chosen the source coordinate system so as to set the polarization angle $\psi$ to $\pi/2$.  Combining  (34) - (36), we can write one signal in terms of the others:
\begin{equation}
h_{3}(t + \Delta t_{13}) = A(\theta,\phi)h_{1}(t) + B(\theta,\phi)h_{2}(t + \Delta t_{12}),
\end{equation} 
where
\begin{equation}
A = \frac{F_{+,2}F_{\times,3}-F_{\times,2}F_{+,3}}{F_{\times,1}F_{+,2}-F_{+,1}F_{\times,2}}
\end{equation}
and
\begin{equation}
B = \frac{F_{\times,1}F_{+,3}-F_{+,1}F_{\times,3}}{F_{\times,1}F_{+,2}-F_{+,1}F_{\times,2}}.
\end{equation}
Many pipelines produce an estimate of $h_{rss}$ (defined by (4)) to quantify the strain amplitude associated with a burst event.  We can rearrange (37) to include terms of this type:
\begin{equation}
h_{rss,3}^{2} = (A^{2}-AB)h_{rss,1}^{2} + (B^{2}-AB)h_{rss,2}^{2} + ABh_{rss,(1+2)}^{2}
\end{equation}
or
\begin{equation}
h_{rss,3}^{2} = (A^{2}+AB)h_{rss,1}^{2} + (B^{2}+AB)h_{rss,2}^{2} - ABh_{rss,(1-2)}^{2}.
\end{equation}

The last terms in (40) and (41) are coherent and refer to the sum/difference of the signals from interferometers 1 and 2.  Both terms are heavily dependent on accurate arrival time recovery.  The relations (40) and (41) can be used to resolve the source direction degeneracy of the three detector geometry by comparing the relative errors of the true location and its mirror image.  Either test (40) or (41) may be more effective depending on the intererometer network and source.  A simulation was performed to determine which check was more efficient for the H1-L1-V1 network.  Linearly polarized gravitational wave sources were placed on an isotropic grid of 4060 points with four different polarization angles chosen randomly from a uniform distribution between $0$ and $\pi/2$ \footnote{The polarization angles were chosen in this manner to avoid redundancy.  The gravitational wave oscillations from $\psi = 0$ to $\psi = \pi$ are identical to those from $\psi = \pi$ to $\psi = 2\pi$, while only a minus sign differentiates the antenna patterns from $\psi = 0$ to $\psi = \pi/2$ and $\psi = \pi/2$ to $\psi = \pi$.  The minus sign should not affect the $h_{rss}$ values used in the check, so only polarizations between $\psi = 0$ and $\psi = \pi/2$ were considered.} at each point.   Note that the poles and points in the plane of the three detectors were excluded as the checks are irrelevant in those directions.  The antenna patterns and $h_{rss}$ ratios for each trial were calculated and then added to noise to see which test held up better as the accuracy of the $h_{rss}$ estimates deteriorated.  Specifically, each of the three $h_{rss}$ values and the sum/difference term were multiplied by one plus a normally distributed random variable with zero mean and standard deviation equal to a specified fraction of the true $h_{rss}$.  This process was repeated for each trial and the results were fed to the checks (40) and (41) and averaged over all trials to yield an efficiency (i.e. the fraction of times the check was successful) for a given relative uncertainty in $h_{rss}$.  This was repeated for several fractions between 0 and 1, giving the plot shown in Figure 8.  This plot demonstrates that (40) is on average more effective for the chosen network, so (40) was chosen for all subsequent tests.

\section*{IV.  Application to Simulated Data}

  The directional estimator of Section II and the amplitude check of Section III were both applied to LIGO-Virgo simulated data. Reference \cite{lvib} describes in detail how the noise in the three interferometers (H1, L1, V1) was produced by filtering stationary Gaussian noise so as to resemble the expected spectrum at design sensitivity in each detector.  Random phase modulation and sinusoids were used to model resonant sources (lines).  Six types of limearly polarized simulated signals were produced using the GravEn algorithm \cite{stuver}: sine gaussians with Q=15 and central frequencies of 235 (SG235) and 820 (SG820) Hz, gaussians of duration 1 (GAUSS1) and 4 (GAUSS4) ms, and two families of Dimmelmeier-Font-Mueller (DFM) supernova core collapse waveforms\cite{dfm} with parameters A=1, B=2, G=1 and A=2, B=4, G=1.  These waveforms are shown in Figure 9.  The injections were configured as above to produce an isotropic distribution of 3960 points on the sky, with each of these grid points being tested with 4 different polarization angles $(\psi = 0, \pi/8, \pi/4, 3\pi/8)$.  The waveforms were delayed and scaled appropriately and added to the noise of each detector for analysis.  The process was repeated for each of the six waveforms.

  A stand-alone parameter estimation \cite{michele} was used to determine the arrival times, their uncertainties, and the $h_{rss}$ values in each trial.  The parameter estimation processed input data by first applying high pass (100 Hz) and linear predictor (whitening) zero-phase filters.  The noise power spectrum was determined by averaging the spectra of an interval before and an interval after the injected signal.  This power spectrum was subtracted from the spectrum of the interval containing the signal and the result divided by the whitening filter frequency response.  The measured amplitude $h_{rss}^{inband}$ was computed over the band from 100 to 1000 Hz as 

\begin{equation}
h_{rss}^{inband} = {\int_{100 Hz}^{1000 Hz} |\tilde{h}(f)|^{2}\,df},
\end{equation}

\noindent subject to a threshold on the signal-to-noise ratio.  Specifically, the data was broken into $N$ frequency bins and the content in a given bin [$f_{i-1},f_{i}$] ($i=1..N$) was only counted if 

\begin{equation}
\rho_i \equiv \sqrt{4\int_{f_{i-1}}^{f_i} \frac{|\tilde{h}(f)|^{2}}{S(f)}\,df} \ge 9.
\end{equation}

\noindent This SNR threshold was chosen to emphasize components with moderate visibility above the noise floor while reducing spurious excess power.  The estimated signal power spectrum was also used to determine the frequency band that contained the middle 90\% of the signal power.  The original whitened time series was zero-phase bandpassed according to this estimate and its maximum taken as the arrival time.  The arrival time uncertainty was taken to be the Cramer-Rao Lower Bound of the arrival time estimator.

\begin{figure}
\includegraphics[width=0.45\textwidth]{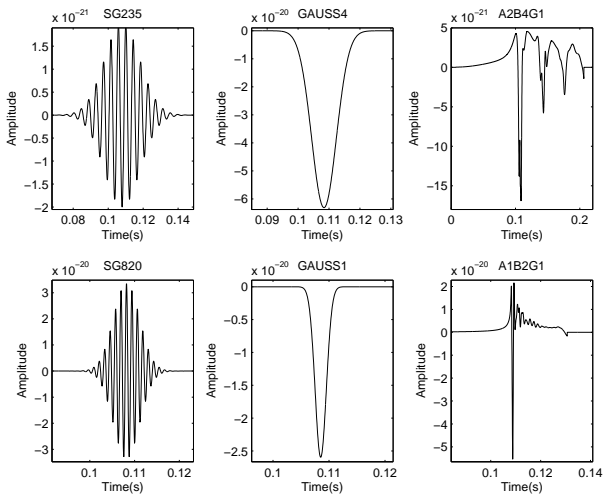}
\caption{Simulated waveforms used for amplitude check testing.}
\end{figure}

The directional estimator consisted of the standard least squares estimator with bias correction as described in Section II.  The bias correction array had directional resolution of 2 degrees azimuthally and 90 bins evenly spaced in the cosine of the polar angle.  Its timing resolution was 0.1 ms in the arrival time uncertainty at each detector.  Trials were conducted with intrinsic inband $h_{rss}$ varying from $10^{-21} Hz^{-1/2}$ to $10^{-18} Hz^{-1/2}$, corresponding to realistic to loud astrophysical sources.  These amplitudes were scaled down by the antenna pattern before being added to the detector noise, on average multiplying the incident $h_{rss}$ by 0.38.  The plot in Figure 10 shows the average angular error for all waveforms as a function of intrinsic inband $h_{rss}$.  This error is defined as

\begin{table}
\begin{tabular}{|c | c | c | c | c | c |}
\hline
Waveform & $10^{-21}$ & $5 \times 10^{-21}$ & $10^{-20}$ & $10^{-19}$ & $10^{-18}$ \\
\hline
SG235 & 0.5415 & 0.8924 & 0.9470 & 0.9947 & 0.9994 \\
\hline
SG820 & 0.0252 & 0.6438 & 0.8076 & 0.9803 & 0.9981 \\
\hline
GAUSS1 & 0.3210 & 0.8215 & 0.9071 & 0.9904 & 0.9993 \\
\hline
GAUSS4 & 0.5537& 0.8936& 0.9458 & 0.9949 & 0.9995 \\
\hline
DFM A1B2G1 & 0.0160 & 0.6416 & 0.8097 & 0.9799 & 0.9979 \\ 
\hline
DFM A2B4G1 & 0.3338 & 0.8238 & 0.9097 & 0.9910 & 0.9992 \\
\hline
\end{tabular}
\caption{Fraction of events with sufficient energy for parameter estimation to estimate parameters.  Note that this fraction is larger than what would be detected by a search algorithm tuned to the same frequency range \cite{lvib} due to inclusion of low energy events that would fall below standard detection thresholds.  The performance of the directional estimator presented here is therefore a conservative estimate due to the inclusion of these low energy events.} 
\end{table}

\begin{equation}
\Delta = \cos^{-1}(\vec{V} \cdot \vec{R}), 
\end{equation}

where $\vec{V}$ is the unit vector in the direction of the estimate and $\vec{R}$ is the unit vector in the direction of the source. The average was taken over all sky locations and polarization angles where a signal was visible to the parameter estimation in all three detectors, and was found to be virtually identical with and without the bias correction.  As can be seen from Figure 11, a few outlying samples caused the average error in each case to be significantly higher than most individual errors.  These outliers were due to incorrect peak time estimates from the parameter estimation.  In most cases the algorithm picked a secondary peak of the signal (see Figure 12), causing the arrival time estimate to be off by a few milliseconds and resulting in a large error in the directional estimate.  The SG820 and DFM A1B2G1 waveforms were particularly susceptible to this, as shot noise becomes firmly established at the higher frequencies where these signals are concentrated.  The lack of impact of the bias correction can be attributed to the relatively small timing uncertainties (approximately 0.1 ms) given by the parameter estimation.  The bias is negligible for such small timing errors, so the correction did not lead to substantial improvement in the trials undertaken.  Even with weaker input signals, the bias correction was seen to be inconsequential as the arrival time uncertainties remained low. It should also be noted that the errors in the arrival time estimates were observed to follow multi-modal distributions in waveforms with multiple peaks, as the most significant errors were a result of the parameter estimation choosing the wrong peak.

\begin{figure}
\includegraphics[width=0.45\textwidth]{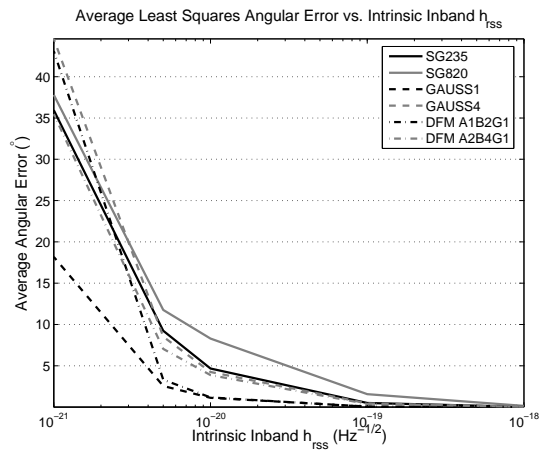}
\caption{Performance of least squares angular estimator on LIGO-Virgo Project Ib simulated waveforms for varying intrinsic inband $h_{rss}$.  In all cases the effect of the bias correction was negligible due to sufficiently small arrival time uncertainties.}
\end{figure}

\begin{figure}
\includegraphics[width=0.45\textwidth]{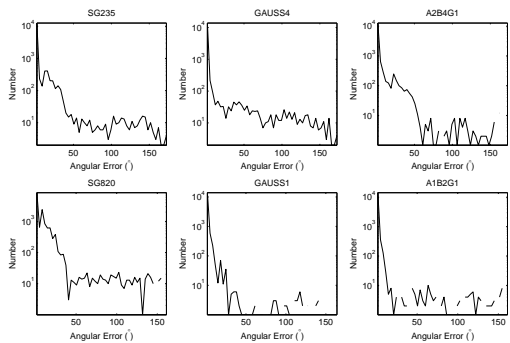}
\caption{Histogram of angular errors for all waveforms with an intrinsic inband $h_{rss}$ of $10^{-20} Hz^{-1/2}$.  Note the logarithmic scale of the y-axis.}

\end{figure}

\begin{figure}
\includegraphics[width=0.45\textwidth]{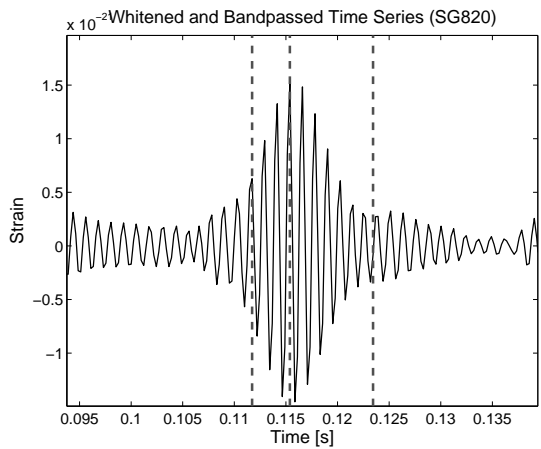}
\caption{Example of difficulty in arrival time estimation.  The parameter estimation picks the largest peak in the whitened, bandpassed data but this may not be the primary peak due to noise.}
\end{figure}

   The observed angular uncertainties can be placed in context through comparison with the classical limit commonly used in radio astronomy\cite{co,kraus,tms}.  The classical limit $\theta$ is given by

\begin{equation}
\theta  \sim \frac{\lambda}{D},
\end{equation}

\noindent where $\lambda$ is the wavelength of the radiation being considered and $D$ is the baseline of the network.  Using the longest such baseline in our network (H1-V1) and assuming a frequency in the middle of our sensitive band (550 Hz) gives a classical limit for our study of $\theta_{GW} \sim 3.8^\circ$.  This differs by less than an order of magnitude from the observed average angular error for all waveforms at an intrinsic inband $h_{rss}$ of $10^{-20} Hz^{1/2}$, which corresponds to a loud astrophysical source (SNR $\sim$ 20).

\begin{figure}
\includegraphics[width=0.45\textwidth]{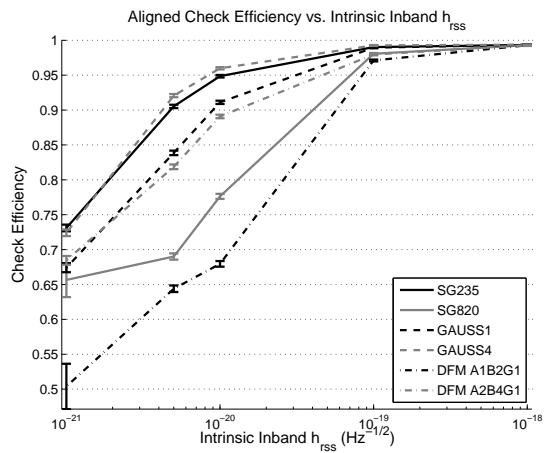}
\caption{Average efficiency of amplitude check (40) as a function of intrinsic in-band $h_{rss}$ for each of the sample waveforms.  The $h_{rss}$ values shown here were scaled down further by the antenna factors before entering the data.  Note that these curves were generated assuming that the time delays between IFOs 1 and 2 were known exactly.}
\end{figure}

\begin{figure}
\includegraphics[width=0.45\textwidth]{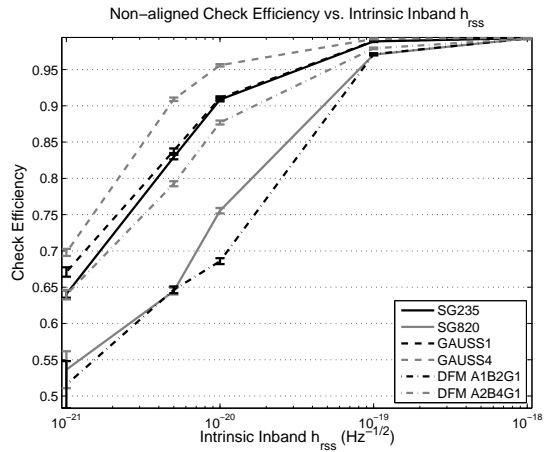}
\caption{Average efficiency of amplitude check (40) as above except with the last term of (40) being determined using parameter estimation peak times.}
\end{figure}

   The same distribution of points, polarization angles, and signal strengths was used to determine the efficiency of the amplitude check (40) as a function of input signal amplitude.  The simulations were again generated in terms of intrinsic inband $h_{rss}$ and scaled down by their antenna factors before being added to the data.  Note that the reported efficiencies are an average over all directions and polarization angles where the signal was detected in all three interferometers.  Two sets of trials were conducted: one where the exact arrival times were assumed to be known for forming the last (coherent) term in (40) and one where the parameter estimation peak times were used for its determination.  The efficiencies were only slightly better with the former ``aligned'' approach, showing that the peak times from the parameter estimation are sufficient to accurately calculate the coherent term.  The efficiencies for the SG820 and DFM A1B2G1 waveforms were slighly below those of the other waveforms due to their power being concentrated at higher frequencies where the interferometer noise is worse.

\section*{V.  Conclusions}

  This work focused on two aspects of coincident source localization.  The first was a least squares analysis that used signal arrival times and their uncertainties to estimate the direction of the source.  This estimator was studied and found to be biased in both the two and three interferometer cases, with the effect increasing with arrival time uncertainty.  The cause of the bias was seen to be twofold as both the allocation of unphysical time delays to their most likely direction and the nonlinear mapping between the time delays and source coordinates contributed to the effect.  The bias will exist for networks with an arbitrary number of detectors, and can be determined numerically through Monte Carlo simulations with a $\chi^2$ minimization routine such as that described in \cite{fabien}.  In each case the bias can be corrected numerically  with the correction being relevant for arrival time uncertainties greater than about 1 ms.  Studies on real data \cite{aps, lvburst} indicate that such uncertainties are probable using conventional detection algorithms.  Our analysis indicates that the bias may also be corrected through the reduction of arrival time errors provided by a thorough parameter estimation such as the one used here\cite{michele}.

   The second piece of coincident analysis described here was an amplitude check, applicable to both linear and multiply polarized signals, that can be used to resolve the source location degeneracy inherent in the three interferometer geometry.  While this analysis is not coincident in a strict sense (one term requires combined data streams), it is extremely simple and so provides a lightweight alternative to more sophisticated methods.  As shown in Section IV, the check is effective for realistic to loud sources, particularly those where the peak time may be recovered accurately.

  One potential application for these source localization techniques would be the generation of skymaps of background and foreground triggers. In this case the amplitude check would improve the efficiency of a distributional test based on sky locations even for cases where its efficiency is only slightly better than 50\%.   We would expect such maps to be isotropic based on current detector sensitivity (i.e. non-gravitational wave triggers dominate) but in the the future such plots may give us a picture of the gravitational wave sky. 

 Methods such as those described above are useful in that they are inexpensive computationally and quite accurate in most cases.  However they may struggle if a source has similar amplitudes in both the + and $\times$ polarizations as well as a delay between the peaks of the waveforms.  Due to the differing antenna responses of the members of an interferometer network, peaks corresponding to the different polarizations may be recorded in different detectors, leading to an inconsistency in arrival time estimates that cannot be accounted for with coincident methods.  The likelihood of this situation is currently under investigation, as is the performance of the algorithms presented here on real data with randomly polarized signals.  

\section*{VI.  Acknowledgements}

LIGO was constructed by the California Institute of Technology and Massachusetts Institute of Technology with funding from the National Science Foundation and operates under cooperative agreement
PHY-0107417. This paper has LIGO Document Number LIGO-P080003-00-Z.

*Available at http://docuserv.ligo.caltech.edu/.


\begin{thebibliography}{99}



\bibitem{fabien} F. Cavalier \emph{et al.}, Phys. Rev. D \textbf{74}, 082004 (2006).

\bibitem{gt} Y. G\"{u}rsel and M. Tinto: Phys. Rev. D 40, 3884 (1989).

\bibitem{scott} E. E. Flanagan, S. A. Hughes, Phys. Rev. D \textbf{57}, 4535 (1998)

\bibitem{sergey} S. Klimenko, S. Mohanty, M. Rakhmanov, G. Mitselmakher, Phys. Rev. D \textbf{72}, 122002 (2005).

\bibitem{patrick} Chatterji \emph{et al.} \emph{Class. Quantum Grav.} gr-qc/0605002 

\bibitem{aps} I. Yakushin:
\emph{Search for Gravitational Wave Bursts in LIGO's S5 Run}, LIGO-G060170 (2006)*.

\bibitem{lvburst} N. Arnaud \emph{et al.}, Phys. Rev. D \textbf{67,} 062004 (2003).

\bibitem{michele} M. Zanolin \emph{et al.}, LIGO-T060226-00-R (2008)*.

\bibitem{ligo3} Abbot B \emph{et al} (LSC) 2006 \emph{Class Quantum Grav.} \textbf{23} S29-39

\bibitem{ykr} I. Yakushin, S. Klimenko, and M. Rakhmanov:
\emph{MDC Frames for S2 Burst Analysis}, LIGO-T040042-00-Z-00-D (2004)*.

\bibitem{dfm}H . Dimmelmeier, J. Font and E. Mueller, Astron. Astrophys. 393 , 523-542 (2002).

\bibitem{lvib}F. Beauville \emph{et al.} 2008 \emph{Class. Quantum Grav.} \textbf{25} 045002 (32pp).

\bibitem{stuver} A. Stuver: \emph{GravEn Simulation Primer}, LIGO-T040020-01-Z (2004)*.

\bibitem{300} K.S. Thorne, in \emph{300 Years of Gravitation}. edited by S. Hawking and W. Israel (Cambridge Universtiy Press, Cambridge, 1987), chap. 9, pp.330-458.

\bibitem{co} Carroll and Ostlie: \emph{An Introduction to Modern Astrophysics} (Addison-Wesley, Reading MA, 1996).

\bibitem{kraus} J. Kraus: \emph{Radio Astronomy, Second Edition} (Cygnus-Quasar Books, Powell OH, 1986).

\bibitem{tms} Thompson, Moran, and Swenson: \emph{Interferometry and Synthesis in Radio Astronomy, Second Edition} (John Wiley \& Sons, Inc., New York, 2001). 

\bibitem{lvibi}F. Beauville \emph{et al.} 2008 \emph{Class. Quantum Grav.} \textbf{25} 045001 (22pp).

\end{thebibliography}
\end{document}